# Effective Gesture Based Framework for Capturing User Input

Pabbathi Sri Charan[1], Saksham Gupta[2], Satvik Agrawal[3], Gadupudi Sahithi Sindhu[1]

[1] MLR Institute of Technology, Dundigal Police Station Road, Hyderabad, Telangana
[2] Chandigarh College of Engineering and Technology, Sector 26, Chandigarh, India,
`dev.sakshamgupta@gmail.com`
[3] Kalinga Institute of Industrial Technology, Patia, Bhubaneswar, Odisha

**Abstract.**
Computers today aren't just confined to laptops and desktops. Mobile gadgets like mobile phones and laptops also make use of it. However, one input device that hasn't changed in the last 50 years is the QWERTY keyboard. Users of virtual keyboards can type on any surface as if it were a keyboard thanks to sensor technology and artificial intelligence. In this research, we use the idea of image processing to create an application for seeing a computer keyboard using a novel framework which can detect hand gestures with precise accuracy while also being sustainable and financially viable. A camera is used to capture keyboard images and finger movements which subsequently acts as a virtual keyboard. In addition, a visible virtual mouse that accepts finger coordinates as input is also described in this study. This system has a direct benefit of reducing peripheral cost, reducing electronics waste generated due to external devices and providing accessibility to people who cannot use the traditional keyboard and mouse.

**Keywords:** Gesture, Artificial Intelligence, Image Classification, Image Detection, Input Devices, Keyboard, Mouse, Virtual Reality, Gesture Controls, OpenCV, CVZone

## 1 Introduction

The importance of human-computer interaction is growing rapidly as computer technology develops, and it has even permeated portable technologies like mobile phones and palmtops. The QWERTY keyboard, a venerable input device, and the most contemporary virtual keyboard technology, however, have not changed in the past fifty years or more [2]. Virtual keyboard has seen use in computer simulation environments and virtual reality systems where access to a traditional keyboard may not be suitable. VR (Virtual Reality) itself offers many tools that enable users to feel and manage virtual objects.

The size of computers has gradually decreased, from "space savers" to "as thin as your hand." The size of discs and other components may have shrunk, but the keyboard has essentially not altered over the years. Techniques using video devices to control mouse movement in the fields of robotics and human computer interaction have been proposed to enable user input in virtual environments [2]. However, these techniques are based on clicking mechanism which requires physical input of some sort. In this study, we are using a virtual keyboard and mouse where we capture finger motions using a camera when operating the mouse and we record finger movements by tapping with the index finger and thumb when using the virtual keyboard. As a result, utilizing a computer is made simpler for people by using a small virtual keyboard and mouse.

## 1.1    Traditional Keyboard Vs Artificial Keyboard

For physical keyboard and mouse-based devices development and deployment take longer. This is due to multiple reasons, including the expense of the physical machinery, manufacturing costs, parts assembly, and labor costs. Traditional keyboards rely on circuits and electronic components which are likely to fail due to hardware damage, water seepage and dust. Physical keyboards and mice also require a power supply to function which further adds to the usage cost. On the other hand, a virtual keyboard does not have such restrictions since there are no physical circuits and the hardware used is very rudimentary. We can control the data by moving a finger on a projector screen, allowing us to write, draw, and text with our fingertips while saving us from having to frequently use the physical keyboard and mouse.

## 1.2    Motivation

The main motivation to build this Artificial keyboard is to reduce the electronic waste and to produce cost effective machines that still perform as efficiently as traditional systems. Accessibility is also a major motivation factor as people with motor diseases like Parkinson's may have difficulty using a traditional keyboard and mouse. Preventing a steep learning curve for users and ensuring adoption in industry was also a consideration while designing this input system. Lastly, this implementation also promotes the use of open-source frameworks like OpenCV and provides a pathway for further improvement AI based input devices. This gesture based artificial keyboard was built by the implementation of Mediapipe and CVZone modules.

## 1.3    Contribution

The main idea behind this study is to develop input devices using gesture-based technology. Minimization of external hardware is cost friendly for the users and better for environment. We have presented a cheap and easy to maintain system that is highly accessible, has minimal environmental impact, is easy to learn and adopt and leads to better user experience.

The rest of the paper is organized as follows. Section 2 discusses related work. Section 3 describes the proposed framework. Section 4 presents the methodology used to create the system. Section 5 describes the results and discussion based on them and finally Section 6 concludes the paper and provides future scope.

## 2    Related Work

Table 1 provides the details of similar research conducted to provide virtual input systems which can replace traditional keyboards and mice. Without the use of a projector or laser light, we used the concept of a camera and image processing to construct a rudimentary keyboard and record the movement of the fingers as they move the mouse.

**Table 1.** Recent research on gesture based virtual input devices

| Publication Year | Authors | Description |
|---|---|---|
| 2020 | Chowdhury et al. [2] | A simple computer-vision based mouse and keyboard relying on hand movements. |
| 2021 | Alhamazani et al. [6] | Hand gesture-based mouse system which can be controlled using a webcam. |
| 2019 | Smeraldi et al. [3] | Detection of hands and fingers using image classification and finding image corners. |
| 2022 | Raschka et al. [5] | A comparative analysis of different python libraries used for image detection and gesture controls. |
| 2022 | Matlani et al. [4] | Using a single camera to act as an input for the mouse gestures and finger detection using edge detection method. |

Several technologies related to virtual keyboard and mouse have been created by numerous researchers in computer science and human computer interaction. But they all employed various strategies. Eckert created keyboard strategies for people with physical disabilities by introducing a novel middleware for mapping movements, collected using a motion sensing camera device. Using an infrared laser module, a keyboard pattern projector, an embedded system, and a single image sensor, Zhang and Yunzhou developed a different strategy. Using image processing techniques including morphological principle and elliptical fitting, they were able to precisely determine each keystroke [20].

In a mouse-related approach, Alsaffar uses fingertip tracking as one method to control the mouse's movement. On the screen, a mouse button click was built so that it would happen whenever a user's hand passed over the area [7][8]. Chu-Feng Lien uses the movement of his fingertip to control the mouse cursor and clicking action. His clicking technique, which was based on visual density, required the user to briefly keep the mouse pointer over the desired location. Paul et al. indicated a thumb clicking event by moving the thumb from the "thumbs-up" posture to a fist.

Gurav used only one conventional camera and one projector and created bare-finger touch contact on typical planar surfaces, such as walls or tables. Just from the 2-D image that the camera has acquired, the touching information of fingertips can be reconstructed [9].

## 3 Proposed Framework

The virtual keyboard and mouse technology creates the keyboard using a camera and hand gestures. This approach is portable like a keyboard, affordable, and simple to use. In earlier systems, the virtual keyboard made use of a projector and camera that recreated the 2D image obtained through touch using only a bare finger [10]. However, employing a projector and laser light limits the usability options and reduced portability. To address these issues, the virtual keyboard was created utilizing only a camera and gesture keyboard. The mechanisms utilized in the design framework are described in the below sections. A flowchart of the framework can be seen in figure 1.

### 3.1 Mouse System

To recognize when the pointer is moving, users will use their finger as the mouse. There will be a camera to record finger movements on the screen in real time. As a result, real-time finger movement will be recognized during image processing. The screen pointer will use these coordinates.

### 3.2 Keyboard System

On a screen, a mock keyboard will be shown. A camera will capture a live feed of fingers pointing towards the camera with the displayed keyboard. As a result, real-time detection of keyboard input will be made possible by image processing. A desktop screen will display those words.

### 3.3 Hand Tracking

Hand tracking is accomplished using OpenCV and MediaPipe. Applications for computer vision use the OpenCV library [1]. MediaPipe performs precise keypoint localization of 3D palm coordinates in the observed hand area using a single-shot palm detection model.

### 3.4 Pose Estimation

A computer vision technique for following the motion of humans and objects is pose estimation which is used to detect and identify the flow of fingers when using the virtual keyboard or mouse [11]. We can contrast various motions and postures based on these crucial considerations and train the movement of mouse based on our specifications.

### 3.8    Corner Rectangle

Corner Rectangle is a library of Python binding created to address issues with computer vision [12]. Any image can have a rectangle drawn on it by using the cv2.rectangle() function. This is used in our system to perform corner detection for keyboard points and fingers.

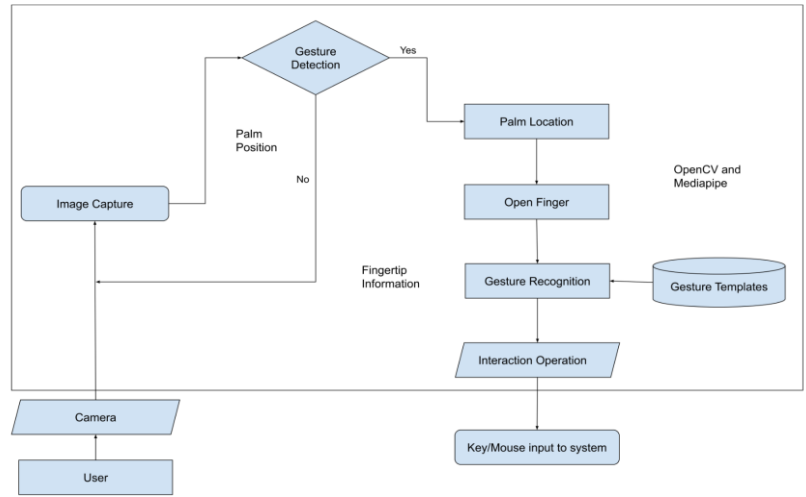

**Fig. 1. Detection Process Flow**

## 4    Methodology

### 4.1    Gestures and Configuration

In our experiment we have used a laptop with a webcam to act as the main system camera. This implementation can be tested on any device that has a front camera and image processing and OpenCV libraries installed [13]. Once configured, the virtual keyboard will appear, and the system is ready to recognize keys.

   The actual click gesture is performed by the user by moving the fingertips inwards followed by an outward motion. Two consecutive frames are captured by the webcam and the gesture is recognized as a click only if these is a difference between the two frames. As soon as any button is clicked correctly, the center of that button will be marked with a random color [14]. The difference between the click area and the center area is calculated so that the button with the coordinates containing the click is recognized and displayed.

   When it comes to the mouse, it also uses a similar pattern with small changes to recognize finger gestures. Move the cursor on the index screen and click with the middle finger [14]. A small smoothing algorithm is used to optimize cursor movement. The Autopsy library helps identify the mouse operations performed and enter them into your computer system.

### 4.2    Detection Process

The first step is to get rid of the background using the camera feed. We obtain an average image feed by using a process called background subtraction that can differentiate between stationary and moving objects and using a reference image, can extract the moving object [19].

Further we utilize contour extraction, which is built in OpenCV, to detect the edges of the fingertips which have a convex hull point. Using this point, missing fingers can be identified by locating the point of maximum deviation [15][16]. A gap-based system is used to detect touch, where is the gap between two fingers reduces past a given threshold, the gesture will be counted as a touch.

For the sake of test automation, self-running demos, and other applications where control of the mouse and keyboard is required, a java class is used to produce native system input events. Robot's main goal is to make automated testing of Java platform implementations easier.

## 5    Results and Discussion

We created a method to get keyboard inputs using gestures and to utilize a live camera to move the mouse cursor and implemented all mouse functions, including left, right, double, and scroll [17]. However, due to varying lighting and background color, the accuracy of the system changed, and maximum accuracy was achieved when the system was placed in a well-lit room with a contrasting background.

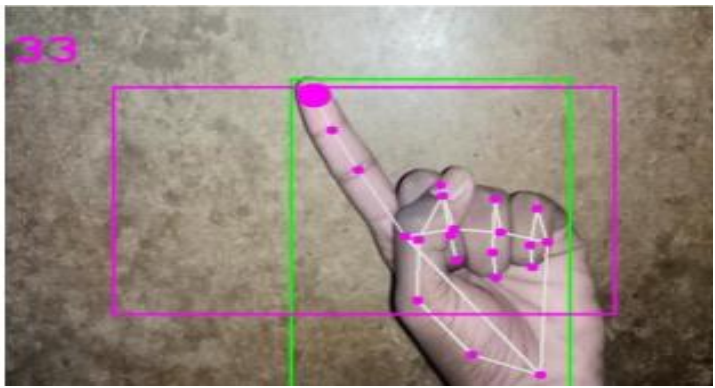

**Fig. 2.**  Moving mouse pointer using figure gestures

Figure 2 demonstrates the mouse and pointer system which can detect the difference between the hand and fingers and are denoted by rectangles of different colors, namely green and pink. The contour lines are projected on extremities and direction of fingers is also represented by independent points representing the overall shape of the gesture.

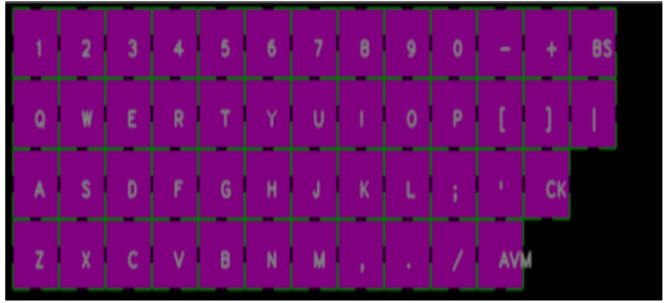

**Fig. 3.** Keyboard displayed on the computer screen

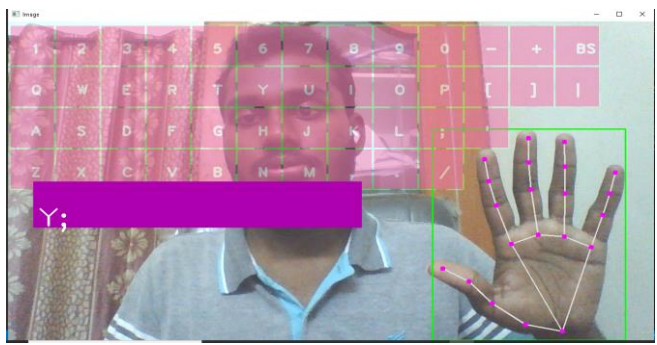

**Fig. 4.** Choosing 'Y' key using a gesture

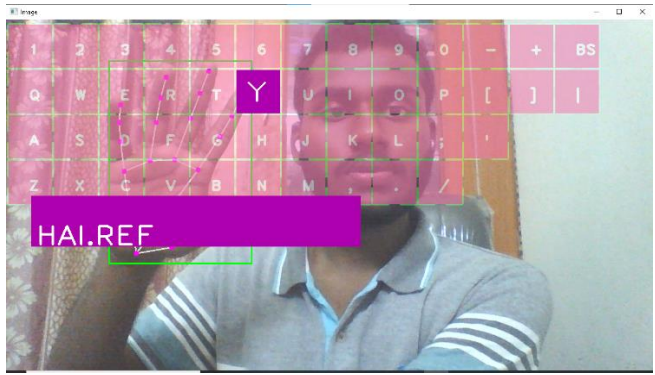

**Fig. 5.** Selecting 'Y' key using a gesture

Figure 3 represents the keyboard system, where the keyboard layout has been presented on the screen. Tapping on the keyboard or using a specific gesture as indicated by figure 4 and figure 5, will find the characters with the closes match and print out the results on the screen.

**Table 2. Accuracy and Detection rate of developed input system**

| Action | Average Detection Rate | Average Accuracy |
|--------|------------------------|------------------|
| Left Click | 95.87% | 99.24% |
| Right Click | 93.89% | 99.38% |
| Double Click | 98.23% | 99.15% |
| Scroll | 91.58% | 99.18% |
| Keypress | 96.91% | 99.92% |
| Point | 98.72% | 99.97% |

To test the accuracy values, we collected over the course of testing with a batch of ten users and three independent machines over fifteen-minute intervals. We logged the entries when an error input was detected. An average of these log values is presented in table 2. We can see that the overall accuracy of the system is well over 99% but the detection rate is unreliable. This difference is due to the disparity in skin tone and the lighting conditions under which the tests were run. However, when the system was able to detect the specific gesture, it was very accurate with the classification of the gesture.

The time to learn how to use the system is minimal and no specific training should be required because the system uses extremely simple gestures based on the concept of point and click.

## 6 Conclusion & Future Scope

Keyboards and mice are essential components of computer systems. With the help of a virtual system we've created, users can interact with a computer without a real keyboard or mouse [18]. This might usher in a new era of computer-human interaction where no direct physical contact is required. It has been demonstrated that using OpenCV to carry out object identification and image processing tasks results in extremely accurate keyboard and mouse task resolution. The maximum accuracy was achieved in the point action at 99.97%.

The fingertip detector module's resistance to changes in lighting and the panel's 3D position estimate for augmented reality of 3D objects will both be the subject of future study on this project. We want to identify projected on-screen touch events in the future by utilizing other graphic elements of the human computer interface, such as character shape and symbolic properties.